\documentclass[twoside]{article}

\evensidemargin\oddsidemargin
\advance \baselineskip 0pt plus 0.1pt minus 0.1pt
\input epsf
\usepackage[T2A]{fontenc}

\usepackage{latexsym}
\usepackage{amsbsy}
\usepackage{amsfonts}
\usepackage{amssymb}

\date{}

\begin{document}

\title{
Charge Density Distributions in Doped Antiferromagnetic Insulators
}
\author{L. S. Isaev, A. P. Protogenov\footnote{e-mail: alprot@appl.sci-nnov.ru} \\
\\
{\fontsize{10pt}{12pt}\selectfont
{\em
Nizhny Novgorod State University, 603950 Nizhny Novgorod
}\/}\\
{\fontsize{10pt}{12pt}\selectfont
{\em
Institute of Applied Physics of the RAS, 603950 Nizhny Novgorod
}}\/}\maketitle
\begin{abstract}
We consider the form of the charge density nano-scale
configurations in underdoped states of planar antiferromagnetic
insulators in the framework of a soft variant of Faddeev-Niemi
model. It is shown that there is such a level of doping and
the temperature range, where charge density distributions in the form
of closed quasi-one-dimensional structures are more preferable.

\end{abstract}

Low-dimensional inhomogeneities in the distribution of charge
\cite{Pan,Hof,T} and spin \cite{BL1,BL2} degrees of freedom at
a moderate level of doping of planar antiferromagnetic insulators
are the subject of active experimental and theoretical
study nowadays. One of the reasons of increased attention to coherent
quasi-one-dimensional charge structures is their existence in the
state, which precedes the high-temperature superconducting phase.
Keeping in mind this key property \cite{PC}, for describing the
mentioned phases we should choose such a model, which will contain
them as limiting cases. In the framework of the mean field theory,
the two-component Ginzburg-Landau model, defined on the complex
projective space $CP^{1}$, satisfies the searched requirements. It
was shown recently \cite{BFN} that there is its exact mapping to
the following extended variant of the ${\bf n}$-field model:
\begin{eqnarray}
  F = \int d^{3}x\left[\frac{1}{4}\rho^{2}\left(\partial_{k}{\bf n}
         \right)^{2}+\left(\partial_{k}\rho \right)^{2}+\frac{1}{16}\rho^{2}
         {\bf c}^{2}+\left(F_{ik}-H_{ik}\right)^{2}+V(\rho, n_{3})\right].
\end{eqnarray}
The free energy in Eq.(1) is defined by the density $\rho^2$, as
well as by the unit vector ${\bf n}$ and the momentum ${\bf
c}={\bf J}/\rho^{2} = 2({\bf j}-4{\bf A})$. They characterize the
spatial distributions of spin degrees of freedom with current
${\bf J}$ or without it. The total current contains the
paramagnetic part ${\bf j}=i[\chi_{1}\nabla\chi_{1}^{\ast}-c.c.+(1
\to 2)]$ and the diamagnetic term $-4{\bf A}$. The following
notations were used in Eq.(3):
$F_{ik}=\partial_{i}c_k-\partial_{k}c_i$ and $H_{ik}={\bf
n}\cdot[\partial_{i}{\bf n}\times\partial_{k}{\bf n}] \equiv
\partial_{i}a_{k}-\partial_{k}a_{i}$. The vector $a_i$ has the
sense of the gauge potential which parametrizes the strength of
the internal magnetic field. Provided the electron spin and charge
are transferred from one of the four sites of some plackets to the
dopant reservoir, the terms with $H_{ik}$ in Eq. (1) characterize
(in the infrared limit) the mean degree of non-collinearity
$<0|{\bf S}_1\cdot\left[{\bf S}_2\times {\bf S}_3\right]|0>$ in
the orientation of three spins remaining in the sites of a
quadratic lattice placket. Therewith the deficit of the charge
density $\rho_{h}^{2}$ relates to the density $\rho^{2}$, which
describes  the distributions of the exchange integral in (1), as
$\rho^{2} + \rho_{h}^{2} = const$. From the long-wavelength point
of view the distribution of the spin density $\rho^{2}$ in the
limited region with the exponential law of decrease at its
boundary (for example, at the distribution in the circle of the
radius $r_{0}$ with the exponential decrease at the length $R\ll
r_{0}$) will be accompanied by the quasi-one-dimensional
distribution of the charge density $\rho_{h}^{2}$ along the
boundary of this region, i.e. along the ring with thickness $R$
and radius $r_{0}$. It is seen from here that studying spatial
configurations of the density field $\rho^{2}$ in planar systems
allows us to find the form of one-dimensional distributions of the
electric charge density with the aid of the above-mentioned
holographic projection.

The multipliers of the first term in the soft variant of the ${\bf n}$-field model (1)
describe the distributions of conjugeted variables -- the spin
stiffness $\;$ $\rho^{2}$ and the square of the inverse length
$(\partial_{k}{\bf n})^{2}$ of the
density screening. It follows from this particular example already, that the
competition of the order parameters $\rho $, ${\bf n}$ and ${\bf c}$ may be the
origin of the existence of phase states with various ordering of strongly
correlated charge and spin degrees of freedom. We enumerate the main limiting
cases of model (1) in inhomogeneous (${\bf n}\not=const$)
phase states as follows:
(i) $\;{\bf c}=0$, $\rho=const$; (ii) $\;{\bf c}=0$, $\rho\not=const$; (iii)
$\;{\bf c}\not=0$, $\rho=const$; (iv) $\;{\bf c}\not=0$, $\rho\not=const$.
In the case of ${\bf n}=const$ and ${\bf c}\not=0$, $\rho\not=const$
the functional (1) is equivalent to the one-component Ginzburg-Landau model.

The model for studying inhomogeneous distributions of spin degrees of freedom in
the form of knots for case (i) was proposed in paper \cite{FN1}. In this rigid
phase $F \geqslant  32\pi^{2}|Q|^{3/4}$ \cite{VK,Ward}, where $Q = 1/(16\pi^{2})\int d^{3}x\,
\varepsilon_{ikl}a_{i}\partial_{k}a_{l}$ is Hopf invariant.
The inhomogeneous superconducting state
(iii) with finite momentum of pairs and universal character of correlations was
recently considered in \cite{PV}. In this phase the gain of the free energy
$F\geqslant 32\pi^{2}|Q|^{3/4}(1 - |L|/|Q|)^{2}$
depends on the index  $L = 1/(16\pi^{2})\int
d^{3}x\,\varepsilon_{ikl}c_{i}\partial_{k}a_{l}$ of mutual linking degree
of one-dimensional configurations, for which the fields ${\bf c}$ and ${\bf a}$ are defined.

The aim of this paper is to study the properties of states outside the
superconducting phase from the second line (ii) of the list of limiting cases,
to which the model (1) leads. In this soft version of the Faddeev-Niemi model \cite{FN2,N}
the functional (1) has the following form
\begin{equation}
  F=\int d^{3}x\left[\frac{1}{4}\rho^{2}\left(\partial_{k}
    {\bf n}\right)^{2} + H_{ik}^{2} + \left(\partial_{k}\rho \right)^{2}
    - b\rho^2+\frac{d}{2}\;\rho^4\right] \, .
 \end{equation}

The state with the broken antiferromagnetic order has a lower 
energy than the "soft" state in which we are interested now. The last one may be 
metastable \cite{PS}. We will consider just such states and compare their 
contribution to the Ginzburg-Landau energy without studying the problems of their 
relaxation, the critical sizes of nuclei of different phases and etc., which are 
of separate interest.

It has been known for a long time (see, for example, \cite{TC}) that
in such a phase state the distributions of the charge density $\rho_{h}^{2}$ have the form of
stripes\footnote{We suppose that the stripe characteristic sizes 
are essentially greater than the lattice scale. In this case, the use of phenomenological 
approach of the Ginzburg-Landau mean field theory is justified.}. 
It is seen from Eq.(2) that the loss in the free energy due to
inhomogeneity may be reduced by developing inhomogeneous configurations of
the field $\rho$ with one-dimensional distribution of the gradient $\partial_k\rho$.
From this point of view it is almost obvious that the quasi-one-dimensional field
configurations $\rho$ in the form of rings give the smallest contribution to the
energy. Whether these spin structures will be two-dimensional with sharp boundaries
or one-dimensional and open, forming stripes, or closed almost one-dimensional
structures in the form of rings, depends also on the parameters of the potential
$V(\rho, n_3)$. In the phase $n_{3}=const$ with neutral spin currents
the answer will depend on the sign of the coefficient $b$ in the potential $V(\rho , n_{3})$.
The positive constant $b$ corresponds to the phase with the broken antiferromagnetic order in (2).
In this case the form of quasi-one-dimensional structures will depend on the result
of the competition between the first and the fourth terms in Eq. (2)
(see also \cite{AN}). We will show that in this phase, far from $T_c$
charge structures with the open ends are preferable, while in the case $T\to T_c$ we should
prefer rings. The first experimental indication of the existence of charge structures
in the form of rings in the underdoped phase of planar systems is in the recent paper \cite{HT}.
In a certain sense, the tunnel microscope in this experiment \cite{HT}
collects data from a two-dimensional slice of knots \cite{FN2}.

Let us find the contribution to the free energy (2) from quasi-one-dimensional
distributions of the density $\rho^2$ in the form of a ring and a stripe and
compare the results of the calculations with the experimental data. We choose the
following trial functions for the configurations of the field $\rho$ in the form of
a ring and a stripe:
\begin{equation}
  \rho=\rho_0\; e^{-(r-r_0)^2/2R^2} \, ,
 \end{equation}
  \begin{equation}
 \rho=\rho_0 exp \left[-x^2/2L_{x}^{2}\right] \times \left\{
 \begin{array}{cc}
 1, & |y|\leqslant L_{y} \, , \\
 exp \left[-(|y|-L_y)^2/2L_{x}^{2}\right], & |y|>L_y \, .
 \end{array}\right.
 \end{equation}
Here $\rho_0=\sqrt{b/d}$, $r_0$ is the ring radius, $R$ is its width,
$2L_y=2\pi r_0$ is the stripe length, $L_x = R$ is its width.
Since configurations (3) and (4) do not depend on the third
coordinate, we shall assume that along this coordinate the size is limited by the length $L_z$,
and also that $R\,<\,r_0$.
The calculation of the energy (2) with the aid of (3) and (4) yields the
following result for the contribution to the free energy from the spin degrees
of freedom distributed over the ring $F_r$ and the stripe $F_{xy}$:
\begin{equation}
  F_r=\pi\rho^2_0 L_z\frac{\bar{r}_0}{R}\left ( 1+
      \frac{R^2}{\xi^2}\right) \, ,
 \end{equation}
 \begin{equation}
  F_{xy}=\pi\rho^2_0 L_z\frac{\bar{r}_0}{R}\left ( 1+
         \frac{R^2}{\xi^2}+\frac{R}{\bar{r}_0}+
         \left( n_0-\frac{3}{4}b \right)\frac{R^3}{\bar{r}_0}
         \right) \, .
 \end{equation}
Here $\bar{r}_0=\sqrt{\pi}r_0$, $1/\xi^2=2\left [ n_0-(1-1/\sqrt{8})b \right]$,
$n_0$ is a certain characteristic value of the "multiplier"
$\;(\partial_k{\bf n})^2$ in (2),
which is of the order $c_1 R^{-2}$, whereas $b=c_2 R^{-2}\delta T$,
where $c_{i} \sim 1$, $\delta T = (T_{c} - T)/T_{c}$.
In these equations we omitted the term $H_{ik}^{2}$ from Eq. (2) since we consider
that it is the same for the both types of distributions.

At $R \ll r_0$ the optimal width of the charge structures $R = \xi$ and the difference of free
energies $\Delta F=F_{xy}-F_r$ in units $\pi\rho^2_0 L_z$ equals
$\Delta F = 1 + c_1 - (3/4)c_2 \delta T$. It follows from this equation that in the case of
$4(1+c_1)/3c_2 < 1$
in the temperature range $\left [ 1 - 4(1+c_1)/3c_2 \right]T_{c}\,<\,T\,<\,T_c$,
which is contigious with the critical temperature $T_c$
of the transition to the state of the spin pseudo-gap, the contribution to the
free energy from the rings is smaller than from the stripes (see Fig. 1).
\begin{figure}
\begin{center}
\epsfxsize=80 mm \epsfbox{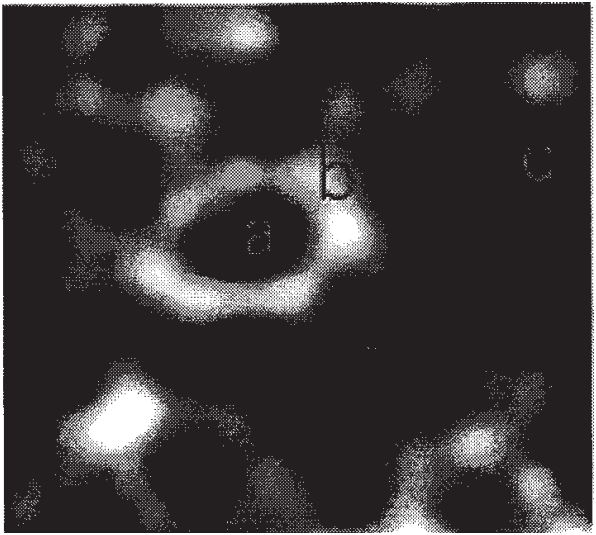}
\end{center}
\caption{
Schematic representation of closed (b) and open (c) quasi-one-dimensional
structures of the charge density (see. \cite{HT}) around antiferromagnetic dielectric
nano-clusters (a).
}
\end{figure}
In the range of temperatures $T\,<\,T_c\left [ 1 - 4(1+c_1)/3c_2 \right]$
the stripes are the main configurations. As is known, one may reach $T_c$,
keeping the temperature constant, by increasing the level of doping.
Under these experimental conditions studied in the recent paper \cite{HT}, it was reported for the
first time that ring-shaped charge structures in underdoped phase
states were observed.

Let us make some remarks concerning the distribution of the spin density $\rho^2$
in the disk, surrounded by the ring charge distribution.
All result were obtained under the assumptiom that
the spin disorder arose only in the region directly adjacent to
the disk edge. Therefore, there was the antiferromagnetic phase inside it.
If the antiferromagnetic order is broken everywhere in the disk, it is
necessary to consider the corresponding distribution of the density $\rho^2$
in the form of a disk with further comparison of its contribution with $F_r$.
When considering the contribution of the density distributions $\rho ^{2}$
to the free energy in the form of a
disk, we get a double gain due to the existence only one edge and
have a loss due to the area. The
calculation shows that at small $R/r_0$ the distributions in the form of rings
appear to be more preferable.

Let us consider the dependence of the critical temperature $T_c$ on the level of
doping. With this aim, we use the relation of $F_r$ to $F_{xy}$ in the following form:
\begin{equation}
  \frac{F_r}{F_{xy}}=\frac{1}{1+B\,R/\bar{r_0}},
 \end{equation}
where
\begin{displaymath}
  B=\frac{1}{2}\biggl [ 1+\frac{1}{2}\frac{n_0-3/4\,b}
    {n_0-1/\sqrt{8}\,b}\biggr]=\frac{3}{4}\;\;
    \frac{n_0-0.68\,b}{n_0-0.65\,b}.
 \end{displaymath}
One can see that configurations in the form of stripes are more
preferable in the range $0.65\,<\,n_0/b\,<\,0.68$
where $F_{xy}\,<\,F_{r}$.
Normalizing the density $\rho^2_0$  to the
particle number $N$ by the condition $N=2\pi\rho^2_0 \xi L_{z} r_{0}m$,
we get the relation between the
parameters $n_0$ and $b$ in the form $x=b/\sqrt{n_0-\alpha b}$.
Here $x=Nd/\sqrt{2}m\pi L_z r_0$ and $\alpha=0.65$.
Thus, for the bounds of the above-considered range where $n_0\sim b$, we
have $T(x)=T_c (1-A\,x^2)$ with a
certain constant $A$. Therefore, inside the region belonging to the phase state with
the broken antiferromagnetic order, there exists a narrower region, located
between the parabolas $T(x)$, where the charge structures have the form of stripes.

The analysis of the inhomogeneous superconducting state
$F_{ik}\not=0,\,\rho\not=const$ is an open problem at present.
Since $(\partial_{k}\rho)^{2}\not=0$, the additional
contribution decreases the gain in the free energy \cite{PV} in
the superconducting phase. The superconducting state from this
point of view will be studied more thoroughly in a separate paper.
Here we mention only that the superconducting current with the
amplitude ${\bf c}_{0}$, flowing along the rings (3), yields the
term of the order ${\bf c}_{0}^{2}R^{2}$ in the multiplier $(1+
{R^2}/{\xi^2})$  of Eq. (5). This explains in particular why the
superconducting region on the phase diagram "temperature - doping
level" $\,$ is shifted to the line $\delta T(x)=0$ of the transition to
the state with the spin pseudo-gap. Indeed, in this case
$V_{eff}(\rho ,
 n_{3}) = -b_{eff}\rho^{2} + (d/2)\rho^{4}$ with $b_{eff} = b - (n_{0} + {\bf c}_{0}^{2}) = (const/R^{2})\delta T$. We see that the finite value of the momentum ${\bf c}$
of superconducting pairs decreases $\delta T$.

In conclusion, we have compared the contributions to the free energy from the
charge density distributions with open and closed ends and found the regions
of the existence of both types of configurations. The obtained results support
the conjecture of the paper \cite{AN} and explains the results of the
experiment described in \cite{HT}.

We are grateful to S. Davis, L.D. Faddeev, S.M. Girvin, E.A. Kuznetsov, B. Lake, A.I. Larkin,
A.G. Litvak, C. Renner, H. Takagi, V.A. Verbus, G.E. Volovik, Y.-S. Wu for advices and useful
discussions. This paper was supported in part by the grant of the Russian
Foundation for Basic Research No. 01-02-17225.

\end{document}